\documentclass[aps,showpacs,prl,twocolumn]{revtex4}
\usepackage{amssymb,amsfonts,amsmath}
\usepackage[usenames]{xcolor}
\usepackage{graphicx}
\usepackage{microtype}
\usepackage{epsfig}
\usepackage{amsmath}
\usepackage{setspace}
\usepackage{layout}
\usepackage{textcomp}
\usepackage{hyperref}
\usepackage{amsfonts}

\begin{document}
\bibliographystyle{apsrev}

\title{Local equilibrium in the Bak--Sneppen model.}
\author{Daniel Fraiman }
\affiliation{Departamento de Matem\'atica y Ciencias, Universidad de San Andr\'es, Buenos Aires, Argentina,}
\affiliation{CONICET, Argentina.}
\email{dfraiman@udesa.edu.ar}

\begin{abstract}
The Bak Sneppen (BS) model is a very simple model that exhibits all the richness of self-organized criticality theory. At the thermodynamic limit, the BS model converges to a situation where all particles have a fitness that is uniformly distributed between a critical value $p_c$ and 1. The $p_c$ value is unknown, as are the variables that influence and determine this value. Here, we study the Bak Sneppen model in the case in which the lowest fitness particle interacts with an arbitrary even number of $m$ nearest neighbors. We show that $p_{c,m}$ verifies a simple local equilibrium relationship. Based on this relationship, we can determine bounds for $p_{c,m}$.   
\end{abstract}
\maketitle

More than 20 years ago, Per Bak and Kim Sneppen introduced one of the most elegant dynamical models of evolution. The model has attracted the attention of numerous physicists, mathematicians and biologists. The Bak Sneppen (BS) evolution model is defined in the following way: there exist $N$ particles, sites or species in a one dimensional ring, and each site $k$ is characterized by a quantity $X_k$, called fitness, which evolves by:
\begin{equation}\label{BS}
X_k(t+1)=\left\{
\begin{array}{lll}
X_k(t) &  &   \mbox{if}\ \ dist(k,\tilde{k}_t)>a \\
U_{k,t} &  & \mbox{if}\ \   dist(k,\tilde{k}_t) \leq a, 
\end{array}
\right.
\end{equation}  
where $\tilde{k}_t=\{k:  \ \  X_k(t) \leq X_j(t)  \ \ \forall j \in \{1,2,\dots, N\}\}$ is the particle with the lowest $X$ value at time $t$. The distance between two particles $i$ and $j$ is $dist(i,j)=min( | i-j |, | i+j-N |),$ just in order to have a ring configuration (periodic boundaries conditions), 
 $U_{k,t}$ are iid random variables with uniform distribution (0,1), and finally $a \in \mathbb{N}$ is the number of neighbors on each side that are interacting with any given particle.  The initial condition is uniform, i.e. $X_k(0)=U_{k,0}$ for all particles.
In \cite{bak93} Bak and Sneppen introduced the model for the case $a=1$ and showed that this extremely simple model, which can be elegantly applied to the evolution of species, exhibits self-organized criticality. Once the system has reached the stationary regime, ``magic'' appears in the model. At the thermodynamic limit, all particles appear with a fitness value that is distributed uniformly between a critical value $p_c$ and 1, and there are avalanches of particle extinction. More formally,  let  $\tilde{X}(t):=X_{\tilde{k}(t)}(t)$  be the lowest fitness value at time $t$ which occurs at site $\tilde{k}(t)$.  
 An avalanche is a succession of events where the lowest fitness value is less than $p_c$. It starts at time $t+1$ if $\tilde{X}(t)\geq p_c$ and $\tilde{X}(t+1)<p_c$, and
 has a duration $\tau$ if $\tilde{X}(t+1)<p_c$, $\tilde{X}(t+2)<p_c$,...,$\tilde{X}(t+\tau)<p_c$, and $\tilde{X}(t+\tau+1)\geq p_c$. 
 This sequence of minimum fitness $\tilde{X}(t+1),\tilde{X}(t+2),...,$ is a dependence sequence which makes calculating the $\tau$ distribution ($P(\tau=s)$) very difficult. Only the first values of the distribution can be easily computed. Nevertheless, it is well known that the mean avalanche duration is infinity, $\langle \tau \rangle=\infty$, due to the power law tail distribution (fingerprint of criticality)~\cite{bak93,bak95,boer,grassberger,paczuski,boettcher,felici}.

 The critical value $p_c$ only depends on the unique parameter of the model, $a$, described in eq.~1 which determines the number of interacting neighbor sites, $m:=2a$ that are updated at each temporal step.  Although there have been large efforts to calculate the value $p_c$, at least for $m=2$, there is no precise result 25 years after the model's introduction. Simulation results show that $p_c(m=2)$ is approximately 0.667~\cite{grassberger,bak95}. In this letter, we study the value $p_c$ for different values of $m$ by presenting a novel local equilibrium property. This property relates $p_c$ with the neighbors of the lowest fitness particle, which allows one to obtain non-trivial bounds for $p_c$.

The strategy adopted here for studying $p_c$ is to focus on the generator of the avalanche, the lowest fitness site, and also its neighbors. The lowest fitness particle is responsible for producing avalanches of very large duration that diffuse very slowly. This behavior was named by the Bak and Sneppen \textit{punctuated equilibrium}. The avalanches can be interpreted as the ``punctuations'' that \textit{maintain} equilibrium.  Clearly, to maintain equilibrium, the number of particles with a fitness value smaller than $p_c$ during the avalanche (known as ``active particles''), must be stable. During the avalanche, there must be an equilibrium where the net flux of particles crossing (up or down) $p_c$ is zero. If this is not true, then the avalanche will grow indefinitely (and no critical behavior will be observed) or disappear very quickly, increasing the fictitious value $p_c$. With this idea in mind, next we write a conserved mass equation with a permeable wall at $X=p_c$.

\begin{figure}[h]
\begin{center} 
\includegraphics[height=12cm,angle=0]{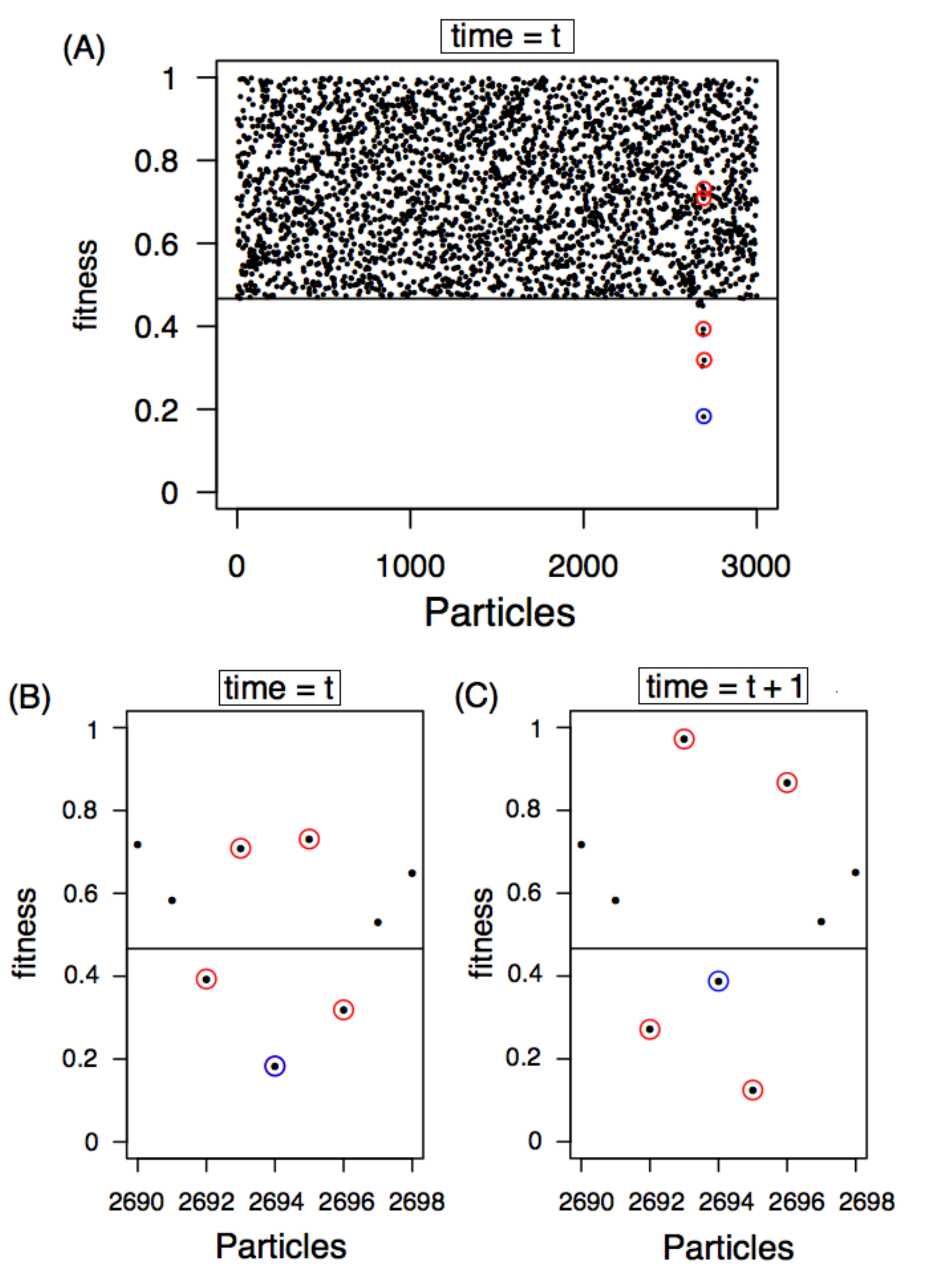}
\end{center}
\vspace{-0.7cm}
\caption{(A) Fitness of the $n=3000$ particles evolved by the Bak--Sneppen model with $m=4$ . (B) Zoom-in of the particles near the lowest fitness particle (2694) at time $t$. The number of neighbors of the lowest particle with fitness below $p_c$ is $S_t=2$.   (C) The same particles shown in panel B at time $t+1$. In this case, the lowest particle is the number 2695 and  $S_{t+1}=1$.   The lowest fitness particle at time $t$ is shown with a blue circle, and the $m$ neighbors of the lowest site are shown with red circles.  } 
\end{figure}

The lowest fitness particle affects the fitness of its $m$ nearest interacting neighbors ($m/2$ each side). Previous to the fitness update, some of these $m$ neighbors are below $p_c$ and some are above. The lowest fitness particle can also have a fitness value larger or smaller than $p_c$ . The latter is the most likely, as it occurs with probability $q= \frac{\langle \tau \rangle}{\langle \tau \rangle+(1-(1-p_c)^{m+1})^{-1}}$, while the former occurs with probability $1-q$. In order for the system to maintain equilibrium, the proportion of particles below $p_c$ must be preserved (i.e. the proportion before the update must be equal to the proportion after the update).  Since updates are uniform, this last proportion is just $p_c$, and the proportion before the update can be easily written using the law of total probability. This preservation gives rise to the following equation:
\begin{equation}
p_c=q\frac{1+\langle S\rangle}{1+m}+(1-q)\frac{0}{1+m},\\
\end{equation}
where $\langle S \rangle$ is the mean number of interacting neighbors that have a fitness value below $p_c$ when the lowest fitness particle is below $p_c$. 
The first right numerator $1+\langle S\rangle$ corresponds to the number of particles that have a fitness below $p_c$ from a total of $1+m$ particles, knowing that the $\tilde{X}$ is below $p_c$. The second numerator term is zero because it corresponds to the case where $\tilde{X}$ is greater or equal to $p_c$ and therefore none of the neighbors can have a fitness value below $p_c$.  Finally, since at the thermodynamic limit $\langle \tau \rangle=\infty$ ($q=1$), we obtain:
\begin{equation}\label{fund}
p_c=\frac{1+\langle S\rangle}{1+m}.
\end{equation}

Another way to think of eq.~\ref{fund} is the following: let us suppose we have a permeable wall at $X=p_c$. At the thermodynamic limit and equilibrium, the lowest fitness particle, $\tilde{k}$, will be (with probability 1) below $p_c$.  Some of the $m$ neighbors of $\tilde{k}$ can be below $p_c$ and some above it. The number of neighbors that are below are equal to $1+\langle S \rangle$, and some of these will cross up the barrier. On average, $(1+\langle S \rangle )(1-p_c)$  will cross up the $p_c$ barrier. On the other hand, above $p_c$ there are $(m-\langle S \rangle)$ particles and on average, $(m-\langle S \rangle)p_c$ will cross down the barrier. If we equal both the number of up cross and down cross particles, imposing that the system maintain equilibrium, we obtain eq. ~\ref{fund}.

 \begin{figure}[h]
\begin{center} 
\includegraphics[height=2.3cm,angle=0]{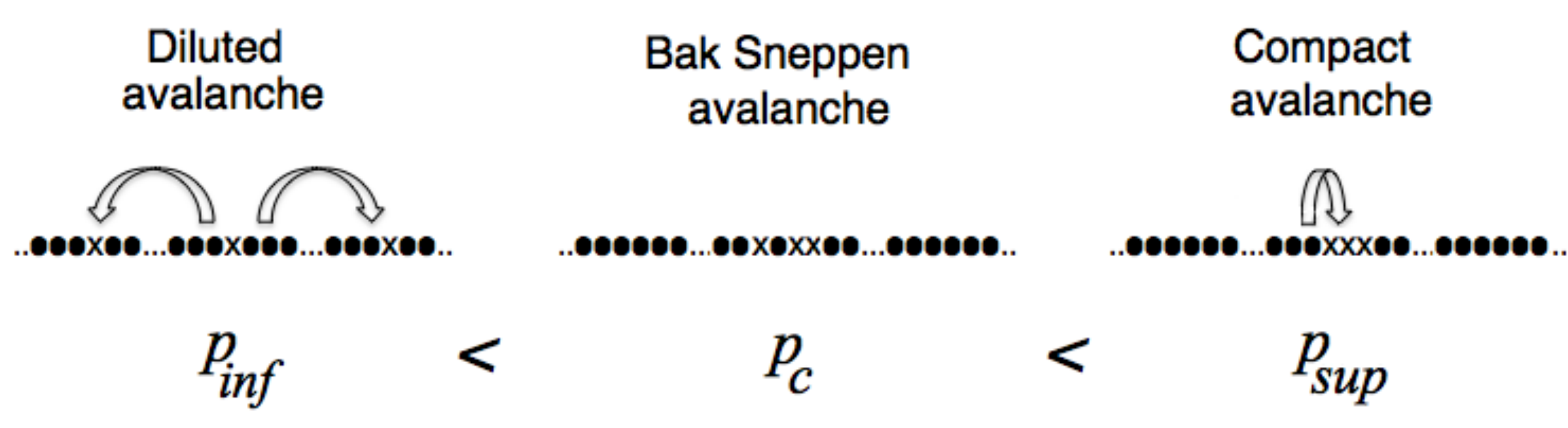}
\end{center}
\vspace{-0.7cm}
\caption{Scheme of the three models studied: random neighbors (left), Bak Sneppen (middle), compact neighbors (right). Crosses represent active particles and points represent inactive ones.}
\end{figure}

Now, based on eq. 2, we show that the critical value of the Bak-Sneppen model can be bounded by the critical value of two different models. These models present a slight modification of the original BS model (Fig. 2). The modification is the following: once the $m+1$ particles of the BS model are updated, we proceed to change the position of the updated particles. In one case, the $m$ updated neighbors are exchanged with other $m$ random particles. This model also presents a critical value called $p_{inf}$, and gives rise to ``diluted'' avalanches (left panel Fig. 2) with $\langle \tau \rangle=\infty$.  At the thermodynamic limit, the neighbors with the lowest fitness particles are always above $p_{inf}$ obtaining in this case $\langle S \rangle$=0, and therefore eq.~\ref{fund} becomes
\begin{equation}\label{cotinf}
p_{inf}=\frac{1}{1+m}.
\end{equation}
In Ref.~\cite{flyvbjerg} the authors obtained the same result, by using mean field theory. Also, consistent with eq.~\ref{cotinf}, in~\cite{meester} found that for the case of one interacting neighbor $m=1$, the critical value is $1/2$.
Note that for this model we can also use a simple branching process argument to obtain the same result. Let $Z(t)$ equal the number of active sites at time $t$, i.e. $Z(t)=\#\{i: X_i(t)<p_{inf} \}$. If we start from a unique ``active particle'', $Z(1)=1$, at the next discrete time point there may be 0,1,2,..., or $m+1$ active particles (born from the first particle that died). This process continues, and each offspring can in turn have anywhere between zero and $m+1$ offspring. Since we are studying the system in equilibrium and at the thermodynamic limit, the probability of selecting the same offspring twice before producing more offspring is zero~\footnote{If the number of particles is infinity, then a particle dies only when it has offspring. There is no chance of being ``killed'' by the offspring of another particle that ``fell'' at the same site.}, then we have a true branching process. It is well known that a branching process is critical if the expected number of offspring is equal to 1. Therefore, in terms of the Bak Sneppen model, we obtain the following equation:  $(m+1)p_{inf}=1$, which is equivalent to eq.~\ref{cotinf}.

Now, we introduce the model that gives rise to an upper bound for $p_c$. In this case, non-consecutive active (below the critical value, now called $p_{sup}$) particles are rearranged so that there are no inactive particles among active ones. In this case, ``trapped'' inactive particles are moved to the border between active and inactive particles. That is why we say that the model generates ``compact'' avalanches (right panel, Fig. 2). In equilibrium and at the thermodynamic limit, particles have a fitness value that is uniformly distributed between $p_{sup}$ and 1 ($U[p_{sup},1]$). Unlike the two previous models, the duration of the avalanches ($\tau$) follows an exponential distribution.    
In this case $\langle S \rangle$ cannot be easily calculated. Nevertheless, we found a superior bound for $p_{sup}$.

We show how to calculate an upper bound for $p_{sup}$ for the case $m=2$.
 Let $Z_t$ be the number of active particles (below $p_{sup}$) at time $t$ with $Z_0=1$, and let $Z=\underset{t\to \infty}{lim}Z_t$ be the stationary version of the process with a mean value $\langle Z \rangle$.  In the stationary condition, any of the $Z=k$ particles that are below $p_c$ can have the lowest fitness, i.e. all have probability $1/k$ of being the lowest fitness particle ($\tilde{k}$). Now, since these $k$ particles are all together (compact) then the number of neighbors of lowest particle, $S$, can be 1 or 2 (or 0 if $k=1$). The value 1 corresponds to $\tilde{k}$ at the edge of active and inactive particles, while the value 2 corresponds to $\tilde{k}$ somewhere ``inside''. Therefore, the mean number of neighbors of lowest particle values verifies
  $$\langle S \rangle=\underset{k\geq 2}{\sum}(\frac{2}{k}1+(1-\frac{2}{k})2)P(Z=k)=2-2\langle \frac{1}{Z} \rangle< 2-2\frac{1}{\langle Z \rangle}. $$   
 Note that if the probability law of $Z$ ($P(Z=k)$) is known, no upper bound for $\langle S \rangle$ is needed.  For $m>2$ the calculation is straightforward and we obtain:
$$p_{sup}=q\frac{1+\langle S\rangle}{1+m}<\frac{1+\langle S\rangle}{1+m}<\frac{1+m-m(2+m)(4\langle Z \rangle)^{-1}}{1+m}.$$
Unfortunately, we do not know how to calculate $\langle Z \rangle$, but we believe it can be calculated since one advantage of this last model is that $Z_t$ can be expressed by a simple birth and death equation.

What else can we say about $\langle S \rangle$ for the Bak-Sneppen model? As mentioned above, $S$ is the number of interacting neighbors that have a fitness value below $p_c$ when the minimum fitness particle is below $p_c$. One can see that since each neighbor is independent, $S$ has a binomial distribution with parameters $m$ and $\tilde{p}$, i.e. $P(S=k)=(^m_k)\tilde{p}^k(1-\tilde{p})^{m-k}$. Then 
\begin{equation}
 \langle S \rangle=m \tilde{p} \quad \quad \mbox{with} \quad \quad  \tilde{p}=\int^{p_c}_0f_S(s)ds,
 \end{equation}
where $f_S(s)$ is the probability density of a randomly selected interacting neighbor of the particle with the lowest fitness level when this value is less than $p_c$.

In order to gain some intuition about $f_S(s)$ we study the Bak-Sneppen at the limit opposite the thermodynamic limit ($N \to \infty$),  that is, the limit of few particles.  Specifically, we study the BS model for a closed system of only $m+1$ interacting particles.  The periodic boundary condition ensures that the $m+1$ particles are always interacting.  In this case, the BS model does not have a critical value $p_c$, but we assume an arbitrary ``fictitious critical value'', $p^*$, just for studying the $S$ distribution when the lowest fitness value is smaller than $p^*$\footnote{In this case, we verify that $p^*=q\frac{1+<S>}{1+m}$ where $q=\frac{\langle \tau_1 \rangle}{\langle \tau_1 \rangle+\langle \tau_0 \rangle}$ with $\langle \tau_1 \rangle =\frac{1}{(1-p^*)^3}$ and $\langle \tau_0 \rangle=\frac{1}{1-(1-p^*)^3}$. Using equation 2, we obtain that  $p*$ can be any value between 0 and 1, which makes sense considering there is no real critical value}. We want to understand how the fitness of neighbors of the minimum fitness particle is distributed when the lowest fitness particle is below a certain value, $p^*$. Let us call $f^{closed}_S(s)$ the corresponding probability density. This probability density can be easily computed from the order statistics distribution. Considering that the fitness $X_i$ of particle $i$ ($i=1,2,..,m+1$) is a uniform (0,1) random variable and that these are independent,  if we call $X_{(1)}$ the lowest value, $X_{(2)}$ the following order statistic, and $X_{(m+1)}$ the maximum fitness value, it is easy to verify that 
\begin{equation}\label{closed}
f_S^{closed}(s)=\frac{1}{mP(X_{(1)}<p^*)}\frac{\partial }{\partial s}\left(\overset{m+1}{\underset{i=2}{\sum}} P(X_{(1)}<p^*, X_{(i)}<s) \right).
\end{equation}
Finally, considering that the joint order statistics pdf for uniform random variables is 
$$f_{X_{(i)},X_{(j)}}(u,v)=N!{u^{i-1} \over (i-1)!}{(v-u)^{j-i-1} \over (j-i-1)!}{(1-v)^{N-j} \over (N-j)!}$$       
 with $0\leq u <v \leq1$ and $N$ is the number of particles (in the closed system $N=m+1$), it is possible to calculate $f_S^{closed}(s)$ from eq.~\ref{closed}.   
 
 For example, for $m=2$ neighbors, we obtain
 \begin{equation*}
f^{closed}_S(s)=\left\{
\begin{array}{lll}
\frac{3s(2-s)}{2(1-(1-p^*)^3)} &  &   \mbox{if}\ \ s<p^* \\
\\
\frac{3p^*(2-p^*)}{2(1-(1-p^*)^3)}  &  & \mbox{if}\ \   s \geq p^*, 
\end{array}
\right.
\end{equation*}
Fig. 3A shows $f_S^{closed}(s)$ for different values of $m=\{2,4,8,20\}$ considering $p^*(m)$ the true $p_c(m)$ observed at the thermodynamic limit. 

\begin{figure}[h]
\begin{center} 
\includegraphics[height=12cm,angle=0]{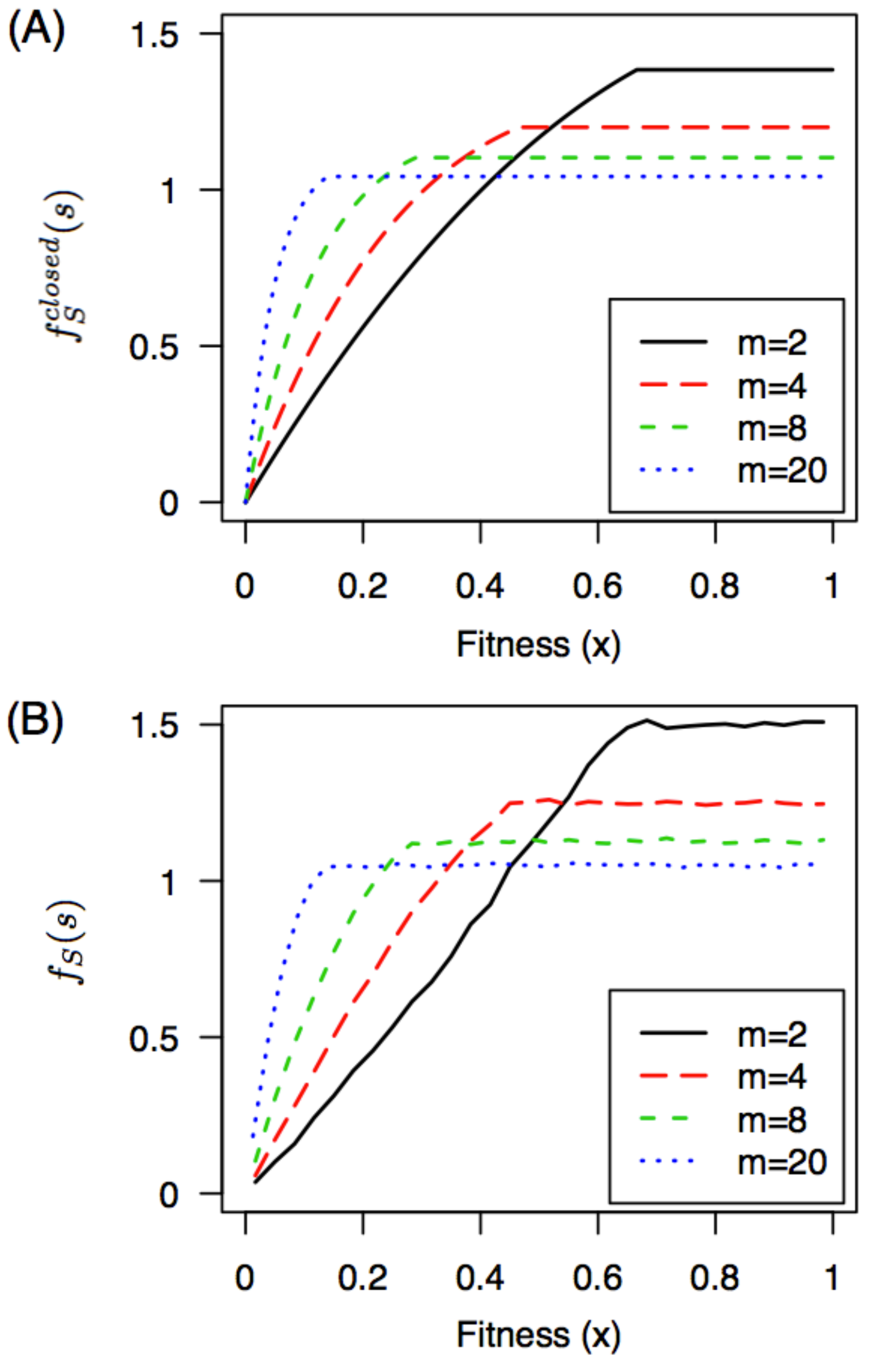}
\end{center}
\vspace{-0.7cm}
\caption{Probability density of the fitness of neighbors of $\tilde{k}$ when $\tilde{k}<p_c(m)$, and when considering: (A) a system of $N=m+1$ particles, (B) a system of $N=4000$ ($\sim$ infinity) particles. $f^{closed}_S(s)$ is calculated from eq. 5.  $f_S(s)$ is estimated by simulations.}
\end{figure}

On the other hand, for large $N$ ($N>>m$) an avalanche can be considered an open system. More particles can be incorporated into the avalanche as time evolves. This is not possible in the closed system where only the $m+1$ fixed particles can be part of the ``avalanche''.  
This difference has an impact on the number of particles from which the minimum fitness is selected. Nevertheless, simulations show that $f_S(s)$ for large $N$ (Fig. 3B) is similar to the one for $N=m+1$ (Fig. 3A). The probability density of $S$, $f_S(s)$, is a smooth function that is partitioned into two sides and can be described as
 \begin{equation}
f_S(s)=\left\{
\begin{array}{lll}
g(s) &  &   \mbox{if}\ \ s<p_c \\
\\
g(p_c) &  & \mbox{if}\ \   s \geq p_c, 
\end{array}
\right.
\end{equation}
with  $g(0)=0$.

 For $m=2$ the probability densities corresponding to $N=m+1$ and to $N=\infty$ present some differences.  But, for $m>2$ densities are similar.  Next, we focus on the case $m>2$. In this case, $f_S(s)$ (or $g$) is a concave function and therefore a lower bound for $p_c$ can be obtained by proposing a linear $g$ function. Under this hypothesis, it is easy to verify that 
 \begin{equation}\label{cotinf2}
\frac{2}{1+m}<p_c,
  \end{equation}
just by using equations 3 and 5.

Finally, in Fig. 4 we show the empirical critical value $p_c$ as a function of $m$ together with the empirical upper bound ($p_{sup}$) and the theoretical lower bounds (eq.~\ref{cotinf} and~\ref{cotinf2}).

\begin{figure}
\begin{center} 
\includegraphics[height=7.0cm,angle=0]{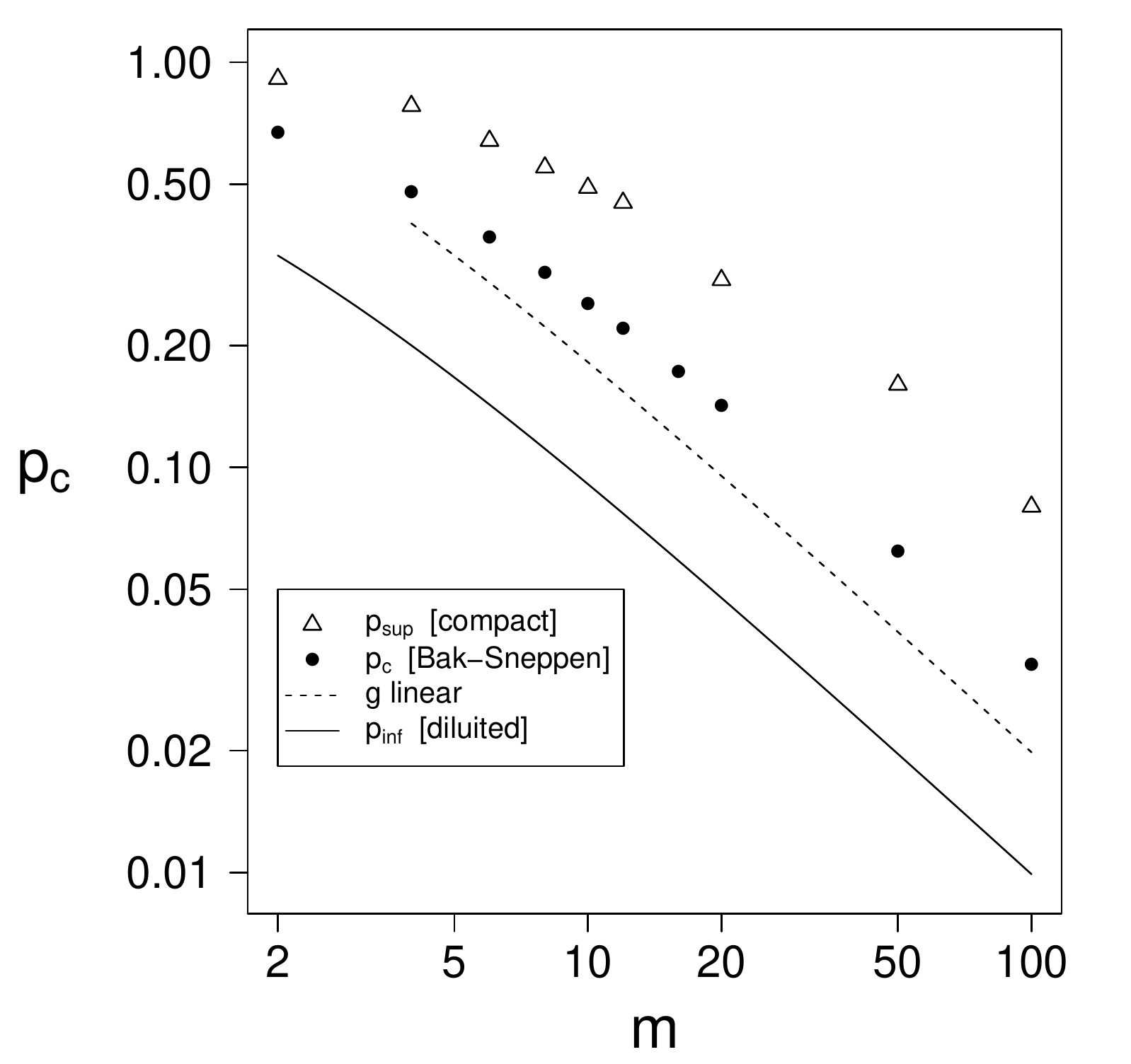}
\end{center}
\vspace{-0.7cm}
 \caption{Critical value of the BS model as a function of $m$. Upper and lower bounds are also represented.} 
 \end{figure}

In summary, in this paper we presented a local equilibrium equation (eq. 2) that allows one to obtain information about the critical value $p_c$ of the Bak-Sneppen model. This equation relates the global selection rule (selecting the lowest fitness particle) with the local cooperative effects (neighbors of the lowest fitness particle are modified). Although we did not study it here, the relationship presented also allows one to study the BS model under different topologies as well as other Bak-Sneppen type models~\cite{dani}.

\end{document}